\documentclass[%
reprint,%
 amssymb, amsmath,%
 aip,apl,%
groupedaddress,%
]{revtex4-1}
\usepackage[dvipdfmx]{graphicx}
\usepackage[italicdiff]{physics}
\usepackage{upgreek}
\usepackage[colorlinks=true,linkcolor=blue]{hyperref}%
\usepackage{comment}
\expandafter\ifx\csname package@font\endcsname\relax\else
 \expandafter\expandafter
 \expandafter\usepackage
 \expandafter\expandafter
 \expandafter{\csname package@font\endcsname}%
\fi
\begin{document}

\title{Detection of nonmagnetic metal thin film using magnetic force microscopy}%

\author{Fujio~Wakaya}%
\email{wakaya@stec.es.osaka-u.ac.jp}
\affiliation{Center for Science and Technology under Extreme Conditions,
Graduate School of Engineering Science,
Osaka University, 
1--3 Machikaneyama, Toyonaka, Osaka 560-8531, Japan}%
\author{Kenta~Oosawa}%
\affiliation{Center for Science and Technology under Extreme Conditions,
Graduate School of Engineering Science,
Osaka University, 
1--3 Machikaneyama, Toyonaka, Osaka 560-8531, Japan}%
\author{Masahiro~Kajiwara}%
\altaffiliation[Present affiliation: ]{Daihatsu Motor Co., Ltd.}
\affiliation{Center for Science and Technology under Extreme Conditions,
Graduate School of Engineering Science,
Osaka University, 
1--3 Machikaneyama, Toyonaka, Osaka 560-8531, Japan}%
\author{Satoshi~Abo}%
\affiliation{Center for Science and Technology under Extreme Conditions,
Graduate School of Engineering Science,
Osaka University, 
1--3 Machikaneyama, Toyonaka, Osaka 560-8531, Japan}%
\author{Mikio~Takai}%
\affiliation{Center for Science and Technology under Extreme Conditions,
Graduate School of Engineering Science,
Osaka University, 
1--3 Machikaneyama, Toyonaka, Osaka 560-8531, Japan}%

\date{Nov. 2, 2018 submitted to Appl. Phys. Lett.; revised on Dec. 2, 2018; 
accepted on Dec. 4, 2018}%

\begin{abstract}
 Magnetic force microscopy (MFM) allows detection of
 stray magnetic fields around magnetic materials and the
two-dimensional visualization of these fields. This paper presents a theoretical analysis of the oscillations of
an MFM tip  above a thin film of
 nonmagnetic metal. The results show good agreement
 with  experimental data obtained by varying the tip height.  
 The phenomenon analyzed here can be applied as a ``metal detector'' at the nanometer scale
 and for contactless measurement of sheet resistivity.  
 The detection sensitivity is obtained as a function of oscillation frequency,
 thus allowing determination of the best frequency for  phase-shift measurement.
 The shift in resonance frequency  due to the presence of a nonmagnetic metal
 is also discussed. \\
 
 \noindent
 Journal reference: Appl. Phys. Lett. \textbf{113}, 261601 (2018).  
 published on-line: Dec. 26, 2018.\\
 DOI: 10.1063/1.5079763
\end{abstract}

\maketitle

Magnetic force microscopy (MFM) is often used for detecting stray magnetic fields near the
surfaces of magnetic materials,\cite{Sarid1994, Rugar1990, Hartmann1999} 
with a magnetized tip oscillating above the material. However, in their investigation of the current-induced magnetic field from  nonmagnetic metal lines,
Tanaka \emph{et al.}\cite{Tanaka2007} reported that
MFM can detect small signals from  nonmagnetic materials (see
  Fig.~4 of Ref.~\onlinecite{Tanaka2007}).
They attributed the observed small signals in the absence of current
to an effect of surface topography, although they did not specify this effect exactly.
Similar small signals have been reported by
Stiller \emph{et al.},\cite{Stiller2017a}
who investigated the current-induced magnetic field using
a ring-shaped nonmagnetic metal line and observed a ring in the MFM image without
any current [see Fig.~3(c) of Ref.~\onlinecite{Stiller2017a}]. However, neither Tanaka \emph{et al.}\cite{Tanaka2007} 
nor  Stiller \emph{et al.}\cite{Stiller2017a} discussed the origin of the
observed small signals in the absence of magnetic materials and of currents.
In the present work,
such MFM signals from a thin film of nonmagnetic metal  are shown to originate from  eddy currents  
 induced in the film by the oscillating MFM tip.
This effect enables us to detect the presence of a nonmagnetic metal even if
it is buried in an insulator, and therefore it could act  as a
``metal detector'' in the nanometer region and  be used for defect inspection in
metal-line layers in integrated circuits.  
Moreover, this effect provides a contactless sheet resistivity measurement method
for metallic materials, including doped semiconductors, where
the dopant concentration could be measured without
the need for the measurement device to come into contact with the probes.  

\begin{figure}[t]
\centering
\includegraphics[width=60mm]{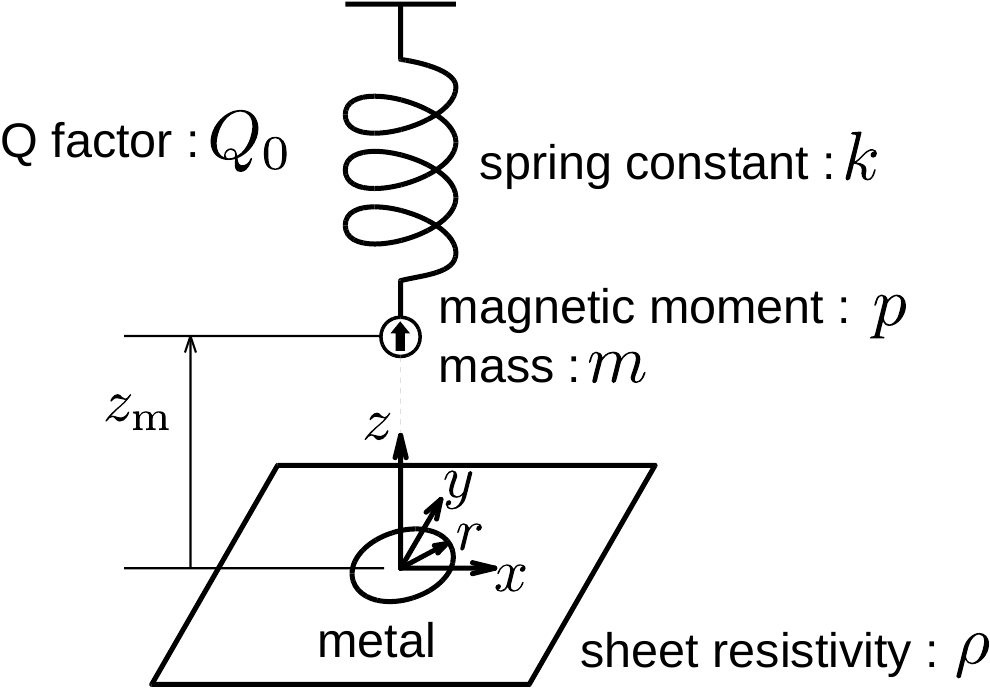}
\caption{Schematic  of the system under consideration. }
\label{fig:1}
\end{figure}
To simplify the real MFM system, the cantilever of the device is modeled as
a spring with spring constant $k$, quality factor $Q_0$, and mass $m$.
Although
the real magnetic dipole moment distributed on the MFM tip surface can be modeled as
a monopole and a dipole moment located at the tip,\cite{Kong1997,Kong1998,Lohau1999}
we assume in this work a single magnetic dipole moment $p$ in the $z$ direction located at the tip end, as is often adopted as a simple model of an MFM tip.\cite{Kong1998}
Figure~\ref{fig:1} shows a schematic of the system.
The equation of motion of the point mass $m$ is
\begin{equation}
m\ddot{z}_\mathrm{m} =
-k\qty[(z_\mathrm{m}-z_\mathrm{m0})-(u-u_0)]
- \gamma\dot{z}_\mathrm{m}
+F_z\, ,
\label{eq:1}
\end{equation}
where
$u$ is the $z$ position of the fixed end of the spring,
$\gamma$ is the dissipation constant of the system,
and $F_z$ is the force in the $z$ direction.
The $z$ positions   $z_\mathrm{m0}$ and $u_0$ are the equilibrium positions of
the point mass and the fixed end, respectively.

The force $F_z$  originates from the motion of the magnetic moment $p$ 
and is obtained as follows within the quasi-static approximation.
The magnetic flux density at a circle of radius $r$ in the $x$--$y$ plane is
\begin{equation}
B_z =
\frac{p}{4\pi}
\frac{2z_\mathrm{m}^2 - r^2}{\qty(z_\mathrm{m}^2+r^2)^{5/2}}.  
\end{equation}
(SI units will be used throughout, but it should be noted that the unit of magnetic moment adopted here is Wb$\cdot$m, rather than A$\cdot$m$^2$.)  
The magnetic flux in a circle of radius $r$ is therefore 
\begin{equation}
\varPhi = \int_0^r B_z 2\pi r' \dd{r'}
= \frac{p}{2}\frac{r^2}{\qty(z_\mathrm{m}^2+r^2)^{3/2}}.
\end{equation}
If the magnetic moment $p$ moves with velocity $\dot{z}_\mathrm{m}$ in the $z$ direction,
then the induced electromotive force, $-\dv*{\varPhi}{t}$, causes an  eddy current $I$
in the metal thin film (which is located in the $x$--$y$ plane) such that
\begin{equation}
\dv{I}{r} =
\frac{3}{4\pi}
\frac{p}{\rho}
\frac{z_\mathrm{m}r}{\qty(z_\mathrm{m}^2+r^2)^{5/2}}\dot{z}_\mathrm{m}\, ,
\label{eq:4}
\end{equation}
where $\rho$ is the sheet resistivity of the metal thin film. 
Interaction between  eddy currents at different radii is ignored
in the present work.  
The eddy current at radius $r$ generates a magnetic field $\dd{H}_z$ in the $z$ direction  at $(0,0,z)$ given by
\begin{equation}
\dv{H_z}{r} =
\frac{3}{8\pi}\frac{p}{\rho}
\frac{z_\mathrm{m}\dot{z}_\mathrm{m}r^3}{\qty(z^2+r^2)^{3/2}\qty(z_\mathrm{m}^2+r^2)^{5/2}}.
\end{equation}
The total magnetic field in the $z$ direction at $(0,0,z)$ is then
\begin{equation}
H_z = \int_0^\infty \dv{H_z}{r}\dd{r} =
\frac{1}{4\pi}\frac{p}{\rho}\frac{\dot{z}_\mathrm{m}}{(z_\mathrm{m}+z)^3}.
\end{equation}
The force $F_z$ in Eq.~(\ref{eq:1}) is therefore
\begin{equation}
F_z = p\dv{H_z}{z}\eval_{z=z_\mathrm{m}}
=-\frac{3}{64\pi}\frac{p^2}{\rho}\frac{\dot{z}_\mathrm{m}}{z_\mathrm{m}^4}\, .
\label{eq:7}
\end{equation}
This equation shows that the direction of the force from the eddy current is
opposite to that of the velocity $\dot{z}_\mathrm{m}$, as expected, which leads to additional dissipation and therefore a reduction in the quality factor, as
discussed later.

Using Eqs.~(\ref{eq:1}) and (\ref{eq:7}), the equation of motion for
the oscillating MFM tip above the metal thin film is
\begin{align}
m\ddot{z}_\mathrm{m}={}&-k\qty[(z_\mathrm{m}-z_\mathrm{m0})-(u-u_0)]\nonumber\\
&- \gamma\dot{z}_\mathrm{m}
- \frac{3}{64\pi}\frac{p^2}{\rho}\frac{\dot{z}_\mathrm{m}}{z_\mathrm{m}^4}.
\label{eq:8}
\end{align}
The excitation of the system is assumed to be
\begin{equation}
u = u_0 + a\cos \omega t,
\end{equation}
where the positive constant $a$ is the amplitude of the exciting oscillation, and
$\omega$ and $t$ are its angular frequency and time, respectively.
In the remainder of  this paper,
because of the  presence of a nonlinear term in Eq.~(\ref{eq:8}),
we adopt the small-amplitude approximation and ignore the higher harmonics.  
With these approximations, 
we can find a solution of Eq.~(\ref{eq:8})
in the form 
\begin{equation}
z_\mathrm{m} = z_\mathrm{m0} + A\cos(\omega t + \phi),
\label{eq:10}
\end{equation}
where the positive constant $A$ is the amplitude of oscillation  of the point mass $m$ and
$\phi$ is the phase shift relative to the exciting oscillation.
The amplitude $A$ and  phase shift $\phi$ can be expressed as 
\begin{equation}
 A^2 = \frac{a^2}{\qty\big[1-\qty(\omega/\omega_0)^2]^2 + \qty(1/Q_I^2)\qty(\omega/\omega_0)^2},
\label{eq:11}
\end{equation}
\begin{equation}
\phi = -\atan\!\qty[\frac{1}{Q_I}\frac{\omega/\omega_0}{1-\qty(\omega/\omega_0)^2}]
-\pi\theta\qty(\frac{\omega}{\omega_0}-1), 
\label{eq:12}
\vspace*{2mm}
\end{equation}
where $\omega_0 \equiv \sqrt{k/m}$ and $\theta(x)$ is the step function.
The modified quality factor $Q_I$ in Eqs.~(\ref{eq:11}) and (\ref{eq:12}) is defined by 
\begin{equation}
 \frac{m\omega_0}{Q_I} \equiv
  \gamma + \frac{3}{64\pi}\frac{p^2}{\rho z_\mathrm{m0}^4},   
\end{equation}
which, owing to the additional dissipation, is smaller than the original quality factor $Q_0$ defined by 
\begin{equation}
\frac{m\omega_0}{Q_0} \equiv 
\gamma. 
\label{eq:14}
\end{equation}
 
%


Using Eq.~(\ref{eq:12}), the phase shift at the resonance frequency
$\omega_\mathrm{r} = \omega_0[1-1/(2Q_0^2)]^{1/2}$
can be calculated as
\begin{equation}
\phi_{\omega_\mathrm{r}} =
-\atan\!\qty[
2Q_0\qty(
1-\frac{1}{2Q_0^2}
)^{1/2}
\qty(
1+Q_0\frac{z_\mathrm{c}^4}{z_\mathrm{m0}^4}
)
],
\label{eq:15}
\end{equation}
where the characteristic tip height $z_\mathrm{c}$ is defined by
\begin{equation}
 z_\mathrm{c}^4 \equiv \frac{3}{64\pi}\frac{\omega_0}{k}\frac{p^2}{\rho}.
 \label{eq:16}
\end{equation}
The phase shift above an insulator (or when the tip is far from the metal surface)
is
\begin{equation}
\phi_{\omega_\mathrm{r}}(\infty) =
-\atan\!\qty[
2Q_0\qty(
1-\frac{1}{2Q_0^2}
)^{1/2}
].
\label{eq:17}
\end{equation}
The phase difference between
above the metal and above an insulator is therefore 
\begin{equation}
 \Delta \phi_{\omega_\mathrm{r}} \equiv
  \phi_{\omega_\mathrm{r}} - \phi_{\omega_\mathrm{r}}(\infty)
\simeq -\frac{1}{2}\frac{z_\mathrm{c}^4}{z_\mathrm{m0}^4},
\label{eq:18}
\end{equation}
where the approximation  $Q_0 \gg 1$ is used.
Equation~(\ref{eq:18}) represents an additional phase delay due to the eddy current in the metal.

\begin{figure}[tb]
\centering
\includegraphics[width=80mm]{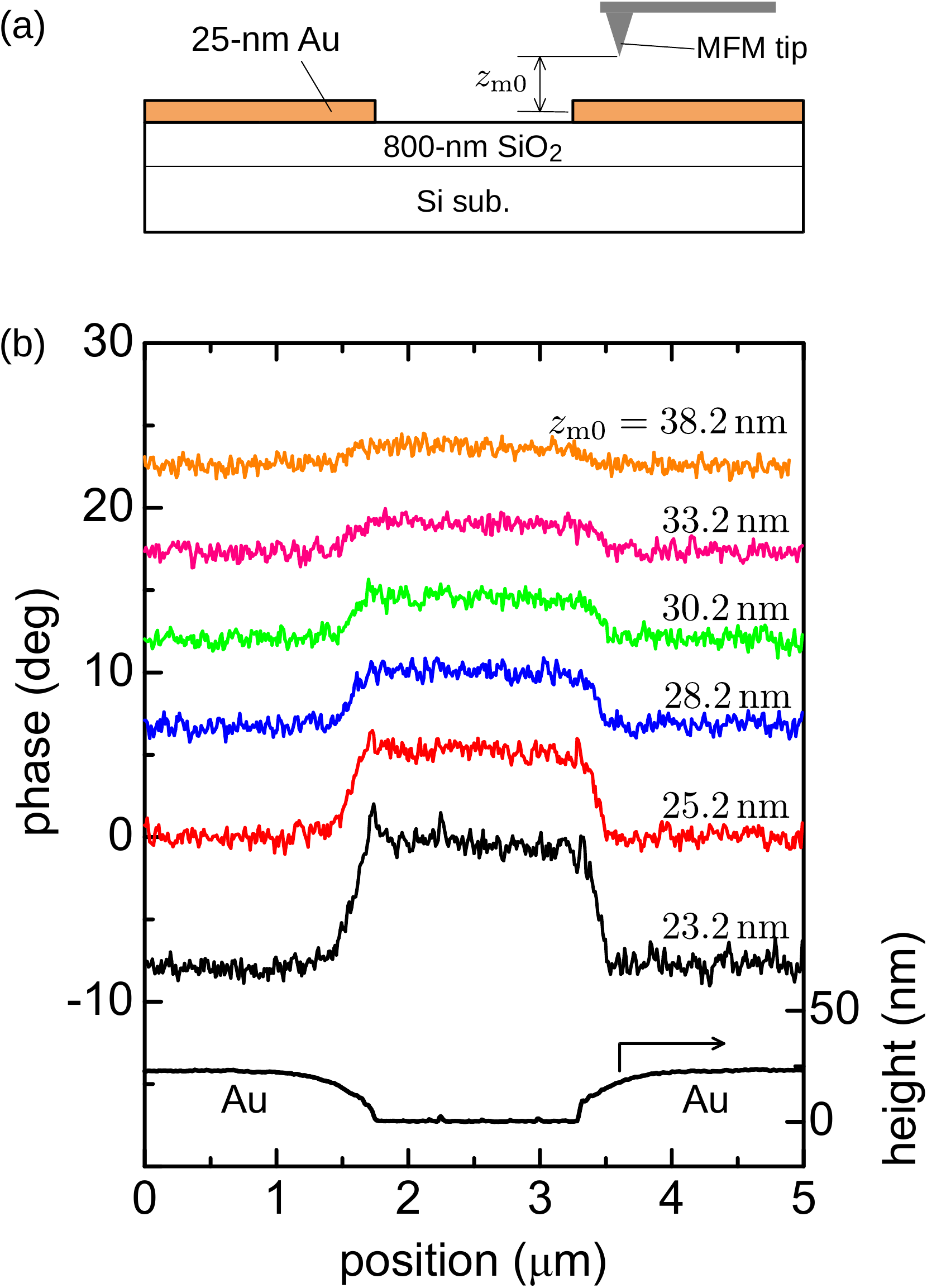}
 \caption{%
 (a) Cross-sectional drawing of the experimental setup. A
 25-nm-thick Au film with 1.5-$\upmu$m gap on  800-nm-thick SiO$_2$ was
 prepared for the experiment.
 (b) Experimental results for phase and topographic height.
$z_\mathrm{m0}$ is measured from the center of the Au film.
The lowest curve shows the topographic height, while all 
the other curves show the phase.
The phase data are shifted vertically for ease of recognition.
}
\label{fig:2}
\end{figure}

\begin{figure}[tb]
\centering
\includegraphics[width=65mm]{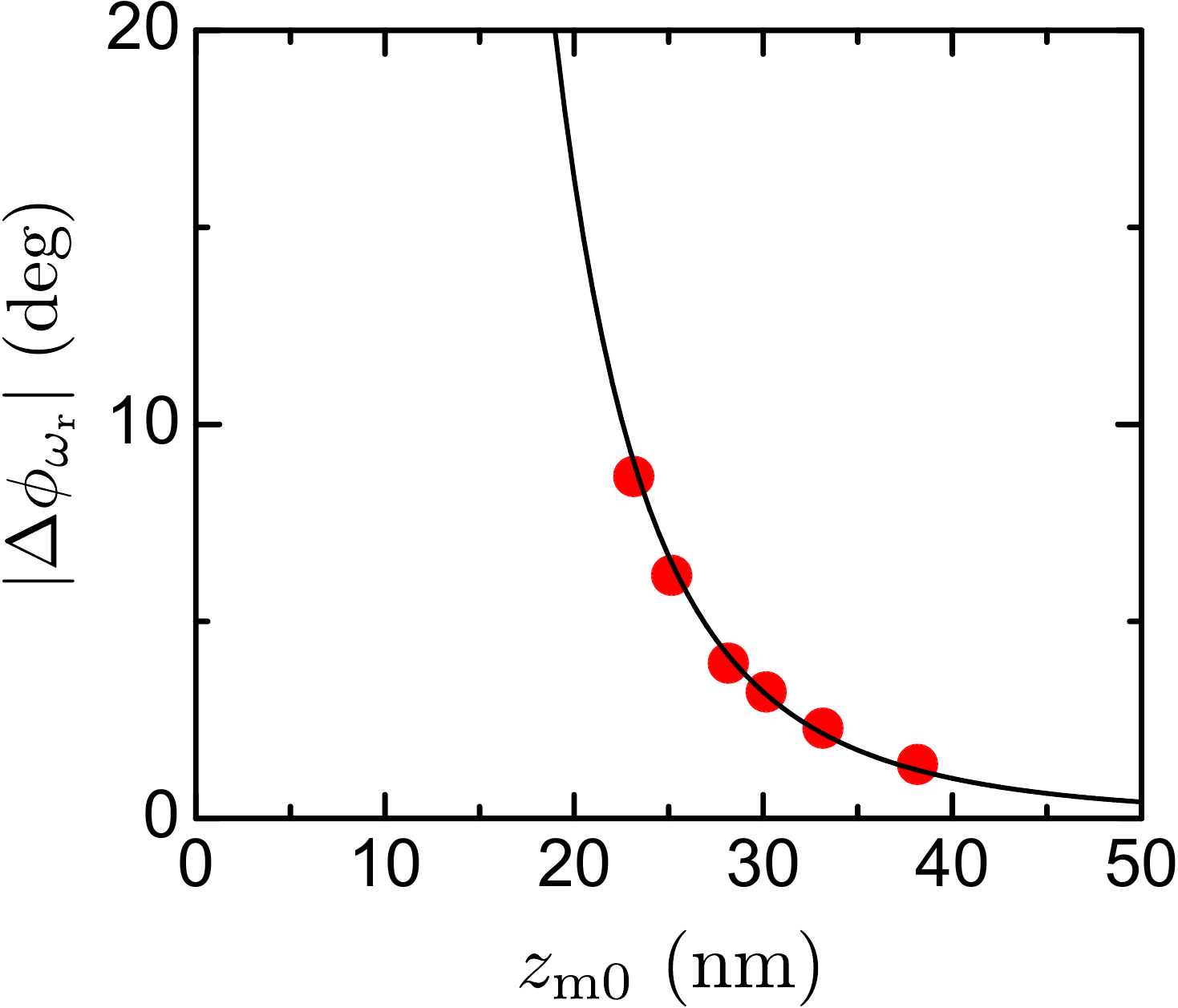}
\caption{%
Observed phase difference as a function of $z_\mathrm{m0}$.
The solid line shows Eq.~(\ref{eq:18}) with $z_\mathrm{c}=17.4$\,nm.
}
\label{fig:3}
\end{figure}

Figure~\ref{fig:2} shows experimental results using a 25-nm-thick Au thin film
with 1.5-$\upmu$m gap deposited on a Si substrate with 800-nm-thick SiO$_2$.
The MFM data were recorded along a 5-$\upmu$m-long line across the 1.5-$\upmu$m gap.
The additional phase delay due to the presence of the nonmagnetic metal thin film
can clearly be seen.
The observed phase difference is
plotted as  a function of $z_\mathrm{m0}$ in Fig.~\ref{fig:3}.  
The experimentally observed phase differences are quite well fitted using
Eq.~(\ref{eq:18}) with $z_\mathrm{c}=17.4$\,nm, which means that
the theory presented above provides a good description of the system of MFM
measurement of a nonmagnetic metal thin film.
We can thus determine the characteristic tip height $z_\mathrm{c}$ 
defined in Eq.~(\ref{eq:16}) by fitting the experimental data with Eq.~(\ref{eq:18}).  
Thus, we can determine the equivalent magnetic moment 
of the MFM tip, $p$, if we already know  the sheet resistivity of the metal, $\rho$, 
or  we can determine $\rho$ if we have already calibrated $p$.  
%
The phase difference between above the metal and above an insulater was not
observed when a tip for normal atomic force microscopy without
magnetic moment was used with similar frequency, dimensions and probe height,
which means that the observed phase difference originated from
the magnetic moment on the tip.  

\begin{figure}[t]
\centering
\vspace*{5mm}
 \includegraphics[width=67mm]{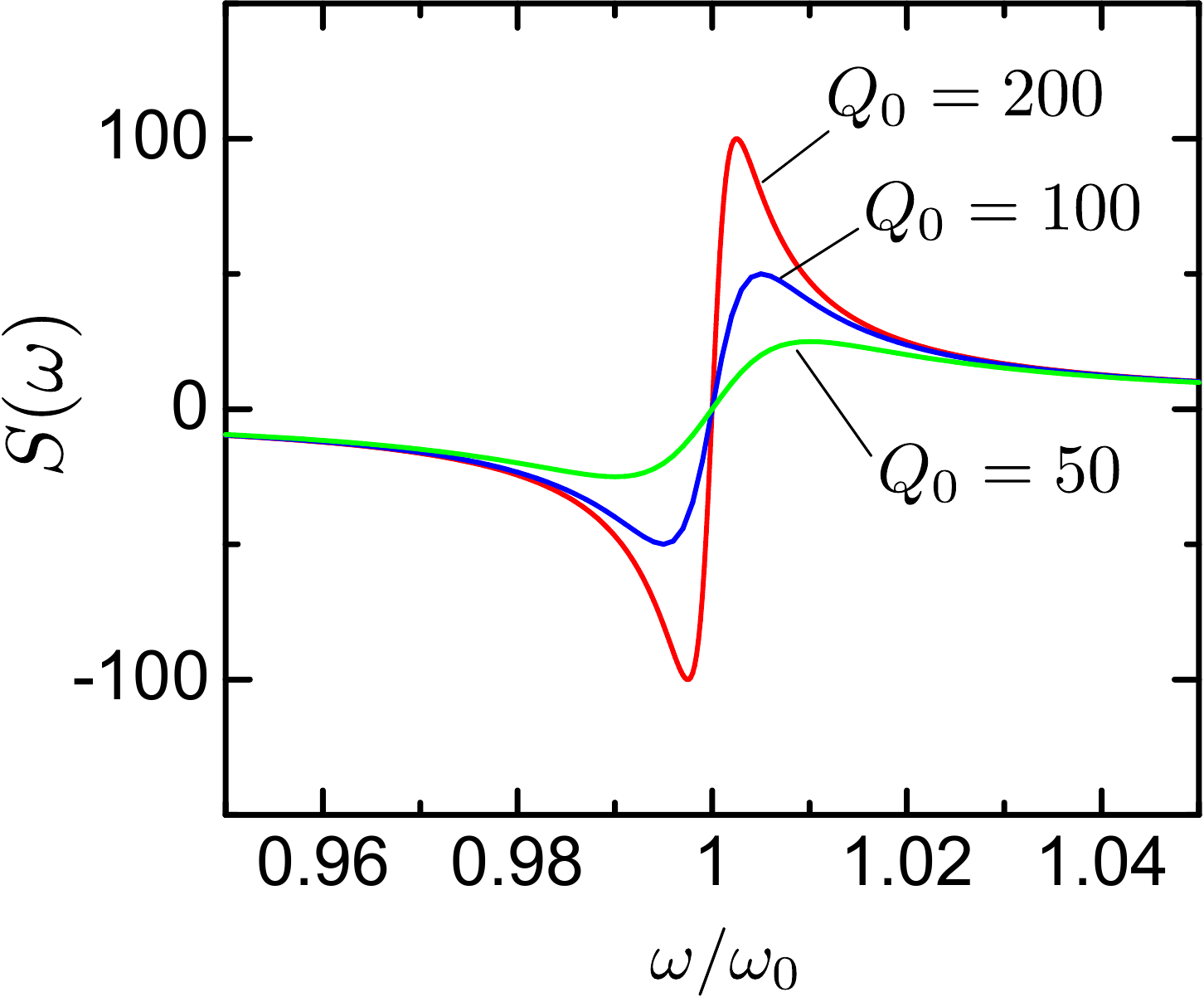}
\caption{%
 Sensitivity of  phase-difference detection for a large tip height 
 and/or a high resistivity 
 as a function of frequency.
}
\label{fig:4}
\end{figure}

The phase difference between above the metal and above an
insulator depends on the oscillation frequency $\omega$,
although this is fixed at $\omega_\mathrm{r}$ in Eqs.~(\ref{eq:15}), (\ref{eq:17}),
and (\ref{eq:18}).
The $\omega$ dependence of the phase shift can be written as
\begin{equation}
 \Delta \phi (\omega) \equiv \phi(\omega, z_\mathrm{m0}) - \phi(\omega, \infty),
\end{equation}
with Eq.~(\ref{eq:12}).
If we define 
\begin{equation}
x \equiv \frac{z_\mathrm{c}^4}{z_\mathrm{m0}^4},
\end{equation}
then $x$ becomes small when the tip height $z_\mathrm{m0}$ becomes large or when
the resistivity $\rho$ becomes large.
This means that
\begin{align}
 S(\omega) &\equiv \dv{\Delta \phi(\omega)}{x}\eval_{x=0} 
 = 
 -\frac{\frac{\omega/\omega_0}{1-(\omega/\omega_0)^2}}
       {1+\frac{1}{Q_0^2}\qty[\frac{\omega/\omega_0}{1-(\omega/\omega_0)^2}]^2}
\end{align}
provides the sensitivity of the phase difference as a function of frequency $\omega$ in the case of 
large tip height and/or a high-resistivity metal thin film.
The sensitivity $S(\omega)$  
is shown in Fig.~\ref{fig:4} for $Q_0=200$, $100$, and $50$.  
As can be seen,
the maximum sensitivity is achieved at
\begin{equation}
 \omega_\mathrm{max}^{\pm} \simeq \omega_0\qty(1\pm \frac{1}{2Q_0}),
\end{equation}
where terms of higher order in $1/Q_0$ are ignored, 
while the sensitivity becomes zero at $\omega = \omega_0$.
At these optimum frequencies, $\omega=\omega_\mathrm{max}^{\pm}$, 
the amplitude of oscillation  becomes $1/\sqrt{2}$ from the maximum, 
which provides a method for finding
the optimum frequencies experimentally.  
The maximum phase difference 
is calculated as 
\begin{equation}
 \Delta \phi(\omega_\mathrm{max}^{\pm}) \simeq
  \pm \frac{1}{2}Q_0 \frac{z_\mathrm{c}^4}{z_\mathrm{m0}^4}, 
\end{equation}
which is a factor of $Q_0$  larger than $\Delta \phi_{\omega_\mathrm{r}}$  
from Eq.~(\ref{eq:18}).  
It is noteworthy that the phase difference  changes  sign 
when the frequency changes across $\omega_0$.

\begin{figure}[t]
\centering
\vspace*{5mm}
\includegraphics[width=55mm]{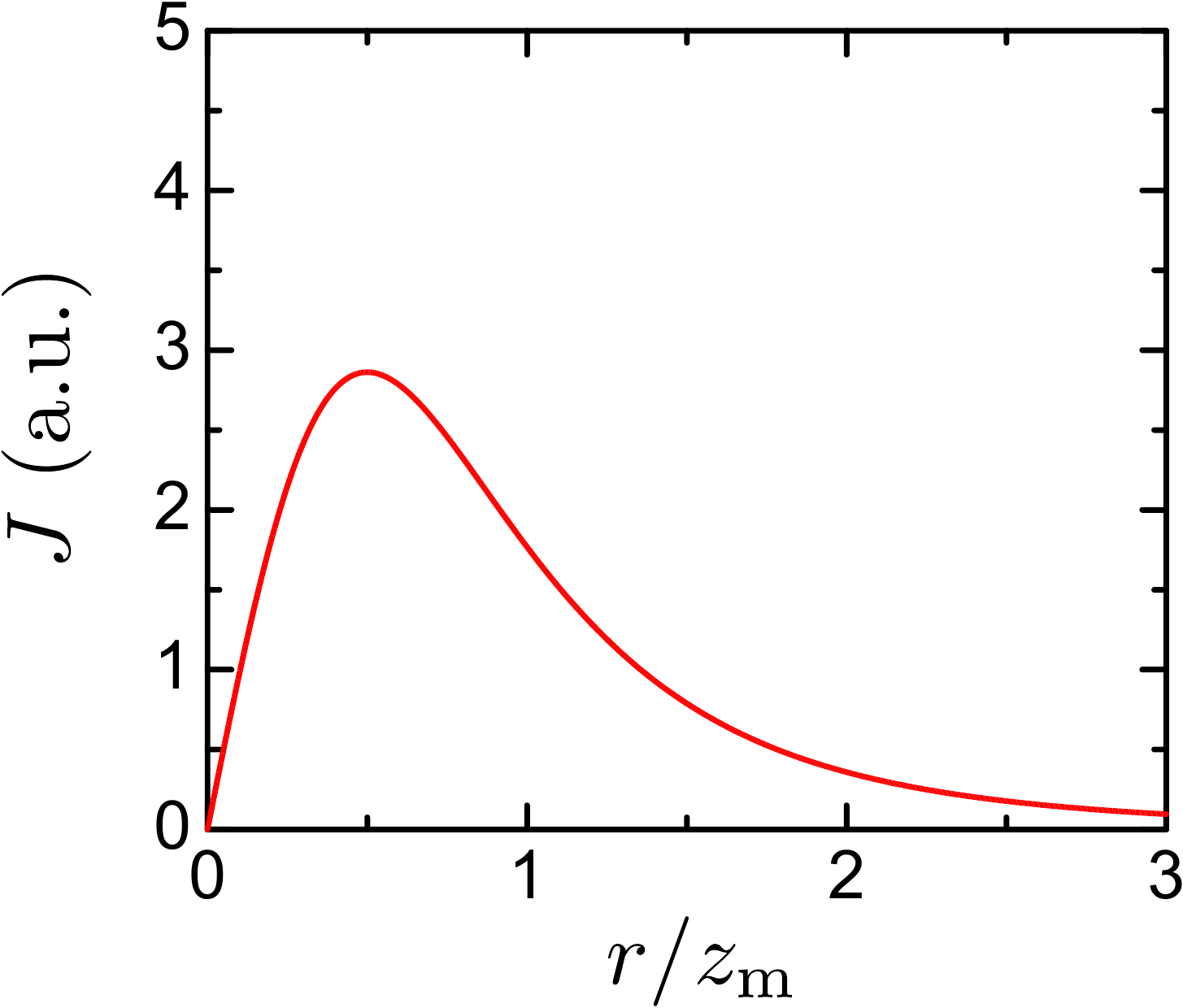}
\caption{%
 Current density of eddy current as a function of radius.  
}
\label{fig:5}
\end{figure}

The spatial resolution achieved when detecting a metal thin film using the phase shift of
the MFM tip oscillation is determined by the distribution of the eddy current
induced by the oscillation.
From Eq.~(\ref{eq:4}), the current density $J$ at
a circle of radius $r$ is 
\begin{equation}
 J \propto \frac{r/z_\mathrm{m}}{\qty[1+(r/z_\mathrm{m})^2]^{5/2}}\, ,
\end{equation}
as shown in Fig.~\ref{fig:5}.  
As can be seen,
most of the eddy current is distributed within a radius
$r\lesssim z_\mathrm{m}$, which means
that the detection of metal regions smaller than the tip height is difficult;
in other words, 
the spatial resolution is roughly limited by the tip height.

Although  phase shifts at a fixed frequency of oscillation  have been discussed so far,
the force from the eddy current induced by  tip oscillation also changes
the resonance frequency through  additional dissipation.
The resonance frequency for $F_z = 0$ is well known to be
$\omega_\mathrm{r} = \omega_0[1-1/(2Q_0^2)]^{1/2}$.
The shift in resonance frequency relative to $\omega_\mathrm{r}$ can be calculated as
\begin{equation}
 \frac{\Delta \omega}{\omega_0} \simeq
  -\frac{1}{4}\qty(\frac{z_\mathrm{c}}{z_\mathrm{m0}})^8,
\end{equation}
where the approximation  
\begin{equation}
 \frac{1}{Q_0} \ll \frac{z_\mathrm{c}^4}{z_\mathrm{m0}^4} \ll 1  
 \vspace*{2mm}
\end{equation}
has been used.  
Such frequency shifts are often utilized in  noncontact atomic force
microscopy with atomic resolution,\cite{Ueyama1995, Katsube2018}
which means that  resonance-frequency detection might have greater sensitivity 
than  phase-shift detection for experimental detection of nonmagnetic metal.

In summary,
theoretical calculations concerning an MFM tip oscillating above a
nonmagnetic metal thin film have provided the phase shift due to the
force from the eddy current induced by the oscillating tip.
The theoretical results are in  good agreement with  experimental observations obtained by
varying the  tip height.
The sensitivity of the phase shift has been shown to be a function of frequency,
which allows determination of the best frequency for  phase-shift measurement.
The shift in resonance frequency  due to  eddy currents has also been discussed.

This work was supported by JSPS KAKENHI Grant No. 18K04937.

%
%
%

%


\begin{thebibliography}{10}%
\makeatletter
\providecommand \@ifxundefined [1]{%
 \@ifx{#1\undefined}
}%
\providecommand \@ifnum [1]{%
 \ifnum #1\expandafter \@firstoftwo
 \else \expandafter \@secondoftwo
 \fi
}%
\providecommand \@ifx [1]{%
 \ifx #1\expandafter \@firstoftwo
 \else \expandafter \@secondoftwo
 \fi
}%
\providecommand \natexlab [1]{#1}%
\providecommand \enquote  [1]{``#1''}%
\providecommand \bibnamefont  [1]{#1}%
\providecommand \bibfnamefont [1]{#1}%
\providecommand \citenamefont [1]{#1}%
\providecommand \href@noop [0]{\@secondoftwo}%
\providecommand \href [0]{\begingroup \@sanitize@url \@href}%
\providecommand \@href[1]{\@@startlink{#1}\@@href}%
\providecommand \@@href[1]{\endgroup#1\@@endlink}%
\providecommand \@sanitize@url [0]{\catcode `\\12\catcode `\$12\catcode
  `\&12\catcode `\#12\catcode `\^12\catcode `\_12\catcode `\%12\relax}%
\providecommand \@@startlink[1]{}%
\providecommand \@@endlink[0]{}%
\providecommand \url  [0]{\begingroup\@sanitize@url \@url }%
\providecommand \@url [1]{\endgroup\@href {#1}{\urlprefix }}%
\providecommand \urlprefix  [0]{URL }%
\providecommand \Eprint [0]{\href }%
\providecommand \doibase [0]{http://dx.doi.org/}%
\providecommand \selectlanguage [0]{\@gobble}%
\providecommand \bibinfo  [0]{\@secondoftwo}%
\providecommand \bibfield  [0]{\@secondoftwo}%
\providecommand \translation [1]{[#1]}%
\providecommand \BibitemOpen [0]{}%
\providecommand \bibitemStop [0]{}%
\providecommand \bibitemNoStop [0]{.\EOS\space}%
\providecommand \EOS [0]{\spacefactor3000\relax}%
\providecommand \BibitemShut  [1]{\csname bibitem#1\endcsname}%
\let\auto@bib@innerbib\@empty
\bibitem [{\citenamefont {Sarid}(1994)}]{Sarid1994}%
  \BibitemOpen
  \bibfield  {author} {\bibinfo {author} {\bibfnamefont {D.}~\bibnamefont
  {Sarid}},\ }\href@noop {} {\emph {\bibinfo {title} {{Scanning Force
  Microscopy}}}},\ \bibinfo {edition} {revised}\ ed.\ (\bibinfo  {publisher}
  {Oxford University Press},\ \bibinfo {address} {Oxford},\ \bibinfo {year}
  {1994})\BibitemShut {NoStop}%
\bibitem [{\citenamefont {Rugar}\ \emph {et~al.}(1990)\citenamefont {Rugar},
  \citenamefont {Mamin}, \citenamefont {Guethner}, \citenamefont {Lambert},
  \citenamefont {Stern}, \citenamefont {McFadyen},\ and\ \citenamefont
  {Yogi}}]{Rugar1990}%
  \BibitemOpen
  \bibfield  {author} {\bibinfo {author} {\bibfnamefont {D.}~\bibnamefont
  {Rugar}}, \bibinfo {author} {\bibfnamefont {H.~J.}\ \bibnamefont {Mamin}},
  \bibinfo {author} {\bibfnamefont {P.}~\bibnamefont {Guethner}}, \bibinfo
  {author} {\bibfnamefont {S.~E.}\ \bibnamefont {Lambert}}, \bibinfo {author}
  {\bibfnamefont {J.~E.}\ \bibnamefont {Stern}}, \bibinfo {author}
  {\bibfnamefont {I.}~\bibnamefont {McFadyen}}, \ and\ \bibinfo {author}
  {\bibfnamefont {T.}~\bibnamefont {Yogi}},\ }\href {\doibase 10.1063/1.346713}
  {\bibfield  {journal} {\bibinfo  {journal} {J. Appl. Phys.}\ }\textbf
  {\bibinfo {volume} {68}},\ \bibinfo {pages} {1169} (\bibinfo {year}
  {1990})}\BibitemShut {NoStop}%
\bibitem [{\citenamefont {Hartmann}(1999)}]{Hartmann1999}%
  \BibitemOpen
  \bibfield  {author} {\bibinfo {author} {\bibfnamefont {U.}~\bibnamefont
  {Hartmann}},\ }\href {\doibase 10.1146/annurev.matsci.29.1.53} {\bibfield
  {journal} {\bibinfo  {journal} {Annu. Rev. Mater. Sci.}\ }\textbf {\bibinfo
  {volume} {29}},\ \bibinfo {pages} {53} (\bibinfo {year} {1999})}\BibitemShut
  {NoStop}%
\bibitem [{\citenamefont {Tanaka}\ \emph {et~al.}(2007)\citenamefont {Tanaka},
  \citenamefont {Mori}, \citenamefont {Yamagiwa}, \citenamefont {Abo},
  \citenamefont {Wakaya},\ and\ \citenamefont {Takai}}]{Tanaka2007}%
  \BibitemOpen
  \bibfield  {author} {\bibinfo {author} {\bibfnamefont {K.}~\bibnamefont
  {Tanaka}}, \bibinfo {author} {\bibfnamefont {Y.}~\bibnamefont {Mori}},
  \bibinfo {author} {\bibfnamefont {H.}~\bibnamefont {Yamagiwa}}, \bibinfo
  {author} {\bibfnamefont {S.}~\bibnamefont {Abo}}, \bibinfo {author}
  {\bibfnamefont {F.}~\bibnamefont {Wakaya}}, \ and\ \bibinfo {author}
  {\bibfnamefont {M.}~\bibnamefont {Takai}},\ }\href {\doibase
  10.1016/j.mee.2007.01.207} {\bibfield  {journal} {\bibinfo  {journal}
  {Microelectron. Eng.}\ }\textbf {\bibinfo {volume} {84}},\ \bibinfo {pages}
  {1416} (\bibinfo {year} {2007})}\BibitemShut {NoStop}%
\bibitem [{\citenamefont {Stiller}\ \emph {et~al.}(2017)\citenamefont
  {Stiller}, \citenamefont {Barzola-Quiquia}, \citenamefont {Esquinazi},
  \citenamefont {Sangiao}, \citenamefont {{De Teresa}}, \citenamefont
  {Meijer},\ and\ \citenamefont {Abel}}]{Stiller2017a}%
  \BibitemOpen
  \bibfield  {author} {\bibinfo {author} {\bibfnamefont {M.}~\bibnamefont
  {Stiller}}, \bibinfo {author} {\bibfnamefont {J.}~\bibnamefont
  {Barzola-Quiquia}}, \bibinfo {author} {\bibfnamefont {P.~D.}\ \bibnamefont
  {Esquinazi}}, \bibinfo {author} {\bibfnamefont {S.}~\bibnamefont {Sangiao}},
  \bibinfo {author} {\bibfnamefont {J.~M.}\ \bibnamefont {{De Teresa}}},
  \bibinfo {author} {\bibfnamefont {J.}~\bibnamefont {Meijer}}, \ and\ \bibinfo
  {author} {\bibfnamefont {B.}~\bibnamefont {Abel}},\ }\href {\doibase
  10.1088/1361-6501/aa925e} {\bibfield  {journal} {\bibinfo  {journal} {Meas.
  Sci. Technol.}\ }\textbf {\bibinfo {volume} {28}},\ \bibinfo {pages} {125401}
  (\bibinfo {year} {2017})}\BibitemShut {NoStop}%
\bibitem [{\citenamefont {Kong}\ and\ \citenamefont {Chou}(1997)}]{Kong1997}%
  \BibitemOpen
  \bibfield  {author} {\bibinfo {author} {\bibfnamefont {L.}~\bibnamefont
  {Kong}}\ and\ \bibinfo {author} {\bibfnamefont {S.~Y.}\ \bibnamefont
  {Chou}},\ }\href {\doibase 10.1063/1.118808} {\bibfield  {journal} {\bibinfo
  {journal} {Appl. Phys. Lett.}\ }\textbf {\bibinfo {volume} {70}},\ \bibinfo
  {pages} {2043} (\bibinfo {year} {1997})}\BibitemShut {NoStop}%
\bibitem [{\citenamefont {Kong}\ and\ \citenamefont {Chou}(1998)}]{Kong1998}%
  \BibitemOpen
  \bibfield  {author} {\bibinfo {author} {\bibfnamefont {L.}~\bibnamefont
  {Kong}}\ and\ \bibinfo {author} {\bibfnamefont {S.~Y.}\ \bibnamefont
  {Chou}},\ }\href {\doibase 10.1063/1.364499} {\bibfield  {journal} {\bibinfo
  {journal} {J. Appl. Phys.}\ }\textbf {\bibinfo {volume} {81}},\ \bibinfo
  {pages} {5026} (\bibinfo {year} {1998})}\BibitemShut {NoStop}%
\bibitem [{\citenamefont {Lohau}\ \emph {et~al.}(1999)\citenamefont {Lohau},
  \citenamefont {Kirsch}, \citenamefont {Carl}, \citenamefont {Dumpich},\ and\
  \citenamefont {Wassermann}}]{Lohau1999}%
  \BibitemOpen
  \bibfield  {author} {\bibinfo {author} {\bibfnamefont {J.}~\bibnamefont
  {Lohau}}, \bibinfo {author} {\bibfnamefont {S.}~\bibnamefont {Kirsch}},
  \bibinfo {author} {\bibfnamefont {A.}~\bibnamefont {Carl}}, \bibinfo {author}
  {\bibfnamefont {G.}~\bibnamefont {Dumpich}}, \ and\ \bibinfo {author}
  {\bibfnamefont {E.~F.}\ \bibnamefont {Wassermann}},\ }\href {\doibase
  10.1063/1.371222} {\bibfield  {journal} {\bibinfo  {journal} {J. Appl.
  Phys.}\ }\textbf {\bibinfo {volume} {86}},\ \bibinfo {pages} {3410} (\bibinfo
  {year} {1999})}\BibitemShut {NoStop}%
\bibitem [{\citenamefont {Ueyama}\ \emph {et~al.}(1995)\citenamefont {Ueyama},
  \citenamefont {Ohta}, \citenamefont {Sugawara},\ and\ \citenamefont
  {Morita}}]{Ueyama1995}%
  \BibitemOpen
  \bibfield  {author} {\bibinfo {author} {\bibfnamefont {H.}~\bibnamefont
  {Ueyama}}, \bibinfo {author} {\bibfnamefont {M.}~\bibnamefont {Ohta}},
  \bibinfo {author} {\bibfnamefont {Y.}~\bibnamefont {Sugawara}}, \ and\
  \bibinfo {author} {\bibfnamefont {S.}~\bibnamefont {Morita}},\ }\href
  {\doibase 10.1143/JJAP.34.L1086} {\bibfield  {journal} {\bibinfo  {journal}
  {Jpn. J. Appl. Phys.}\ }\textbf {\bibinfo {volume} {34}},\ \bibinfo {pages}
  {L1086} (\bibinfo {year} {1995})}\BibitemShut {NoStop}%
\bibitem [{\citenamefont {Katsube}\ and\ \citenamefont
  {Abe}(2018)}]{Katsube2018}%
  \BibitemOpen
  \bibfield  {author} {\bibinfo {author} {\bibfnamefont {D.}~\bibnamefont
  {Katsube}}\ and\ \bibinfo {author} {\bibfnamefont {M.}~\bibnamefont {Abe}},\
  }\href {\doibase 10.1063/1.5037741} {\bibfield  {journal} {\bibinfo
  {journal} {Appl. Phys. Lett.}\ }\textbf {\bibinfo {volume} {113}},\ \bibinfo
  {pages} {031601} (\bibinfo {year} {2018})}\BibitemShut {NoStop}%
\end{thebibliography}
\end{document}